\documentclass[prc,preprint,eqsecnum]{revtex4}
\usepackage{amsmath}
\usepackage{graphicx}
\usepackage{bm}

\begin{document}
\preprint{NT@UW-04-012}

\title{Chiral solitons in nuclei: Electromagnetic form factors}
\author{Jason R. Smith}
\author{Gerald A. Miller}
\affiliation{Department of Physics\\
University of Washington\\Seattle, WA 98195-1560}
\begin{abstract}
We calculate the electromagnetic form factors of a bound proton.
The Chiral Quark-Soliton model provides the quark and antiquark
substructure of the proton, which is embedded in nuclear matter.
This procedure yields significant modifications of the form
factors in the nuclear environment. The sea quarks are almost
completely unaffected, and serve to mitigate the valence quark
effect. In particular, the ratio of the isoscalar electric to the
isovector magnetic form factor decreases by 20\% at $Q^2=1 \text{
GeV}^2$ at nuclear density, and we do not see a strong enhancement
of the magnetic moment.
\end{abstract}
\maketitle

\section{Introduction}
\label{sec:intro}

Recent polarization transfer experiments at TJNAF
\cite{Strauch:2002wu} observed a difference in the electromagnetic
form factors of a proton bound in a Helium nucleus compared to a
free one. This, along with other effects, such as the nuclear EMC
effect \cite{Aubert:1983xm}, seems to suggest the modification of
hadrons in the nuclear medium.

There is extensive work on the medium modifications of
electromagnetic properties of the nucleon in the literature (for
example, see
Refs.~\cite{RuizArriola:ex,Frank:1995pv,Yakhshiev:2002sr,Lu:1998tn}
). This includes effective Lagrangians as well as models that
include the quark substructure of hadrons. While in principle
these effects could be couched in terms of effective field theory
operators, it is our thesis that such results may be more
transparent, physically intuitive or straightforward to calculate
when viewed as a change in the internal structure of the hadrons.

We will use the Chiral-Quark Soliton model (CQS)
\cite{Diakonov:2000pa,Christov:1995vm}, which has a direct
connection to QCD via the Instanton Liquid model, to provide our
subnuclear degrees of freedom. The primary motivation is that this
model includes sea quarks which we have seen to be important in
the nuclear EMC effect \cite{Smith:2003hu}. In that case, the
large medium modification in the valence quark sector is reduced
through the lack of such an effect in the sea (which can be seen
directly in Drell-Yan experiments \cite{Alde:im}). The CQS is
combined with the nuclear medium in a self-consistent quark-meson
coupling calculation as in our previous work \cite{Smith:2003hu},
and the electromagnetic form factors are extracted via the wave
functions of the quarks using the results of
Ref.~\cite{Christov:1995hr}. The overall procedure is similar to
the Quark-Meson Coupling model (QMC) \cite{Lu:1998tn}, which uses
the MIT bag model for the nucleon. The bag model does not include
sea quarks. It is a confining model, whereas the CQS model is not.
Additionally, the QMC model calculation, when coupled with a
Relativistic Distorted Wave Impulse Approximation (RDWIA)
calculation \cite{Udias:1999tm} or a Relativistic
Multiple-Scattering Glauber Approximation (RMSGA) calculation
\cite{Ryckebusch:2003fc,Lava:2004mp}, improves the agreement
between theory the TJNAF data \cite{Strauch:2002wu}. With our
study, we hope to reinforce the interpretation of the medium
effect in terms of quark degrees of freedom, as well as provide an
alternate model when the accuracy of the data is improved.

We begin with a brief description of the CQS model in Section
\ref{sec:model}. In Section \ref{sec:medium}, we motivate and
present our procedure to embed this model in nuclear matter. This
description differs only slightly from that in our previous work
\cite{Smith:2003hu}; it is repeated for completeness.
Subsequently, we describe the numerical methods, and proceed to
the results in Section \ref{sec:results}.

\section{Chiral Quark-Soliton Model}
\label{sec:model}

The CQS model Lagrangian with (anti)quark fields
$\overline{\psi},\psi$, and profile function $\Theta(r)$ is
\cite{Diakonov:2000pa}
\begin{equation}
\mathcal{L} =  \overline{\psi} ( i \partial \!\!\!\!\!\:/\, - M
e^{  i \gamma_{5}\bm{n}\cdot\bm{\tau} \Theta(r) } ) \psi,
\label{eq:lagrangian}
\end{equation}
where $\Theta(r\rightarrow\infty) = 0$ and $\Theta(0) = -\pi$ to
produce a soliton with unit winding number. The quark spectrum
consists of a single bound state and a filled negative energy
Dirac continuum; the vacuum is the filled negative continuum with
$\Theta = 0$. The wave functions in this spectrum provide the
input for the electromagnetic form factors.

We work to leading order in the number of colors ($N_{C}=3$), with
$N_{f}=2$, and in the chiral limit. While the former characterizes
the primary source of theoretical error, one could systematically
expand in $N_{C}$ to calculate corrections. We take the
constituent quark mass to be $M=420\text{ MeV}$, which reproduces,
for example, the $N$-$\Delta$ mass splitting at higher order in
the $N_{C}$ expansion, as well as many electromagnetic properties
\cite{Christov:1995vm,Christov:1995hr}.

The theory contains divergences that must be regulated. We use a
single Pauli-Villars subtraction. The Pauli-Villars mass is
determined by reproducing the measured value of the pion decay
constant, $f_{\pi} = 93 \text{ MeV}$, with the relevant divergent
loop integral regularized using $M_{PV}\simeq 580 \text{ MeV}$.

The nucleon mass is given by a sum of the energy of a single
valence level ($E^{v}$), and the regulated energy of the soliton
($E_{\Theta}$, equal to the sum of energy levels, $E_{n}$, in the
negative Dirac continuum with the sum of the energy levels in the
vacuum, $E_{n}^{(0)}$, subtracted)
\begin{subequations}
\label{eq:mn}
\begin{eqnarray}
M_{N} & = & N_{C} E^{v}+
E_{\Theta}(M)-\frac{M^{2}}{M_{PV}^{2}}E_{\Theta}(M_{PV})\\
E_{\Theta}(M') & = & \sum_{E_{n},E_{n}^{(0)}\leq 0} E_{n} -
E_{n}^{(0)}\Bigg|_{M=M'}.
\end{eqnarray}
\end{subequations}
The field equation for the profile function, which follows from
the Lagrangian (\ref{eq:lagrangian}), is
\begin{equation}
\Theta(r) = \arctan
\frac{\rho_{ps}^{q}(r)}{\rho_{s}^{q}(r)},\label{eq:thetafe}
\end{equation}
where $\rho_{s}^{q} \text{ and } \rho_{ps}^{q}$ are the quark
scalar and pseudoscalar densities, respectively, and are given by
sums of the wave functions of every occupied energy level.

The electromagnetic form factors are also given in terms of the
wave functions, and are derived in Ref.~\cite{Christov:1995hr}.
The formulae are reproduced here, with a Pauli-Villars regulator,
for convenience. To leading order in $N_{C}$, we have only the
isoscalar electric and isovector magnetic form factors
($G_{X}^{T=0,1} = G_{X}^{p} \pm G_{X}^{n}$)
\begin{widetext}
\begin{subequations}\label{eq:GEM}
\begin{eqnarray}
G_{E}^{T=0}(q^{2}) & \stackrel{N_{C}\rightarrow\infty}{=} &
\frac{N_{C}}{3} \int d\bm{r}\: e^{i\bm{q}\cdot\bm{r}} \Bigg\{
\sum_{E_{n}\leq E^{v}} \psi_{n}^{\dag}(\bm{r})\psi_{n}(\bm{r})
-\sum_{E_{n}^{(0)}\leq 0} \psi_{n}^{(0)\dag}(\bm{r})\psi_{n}^{(0)}(\bm{r}) \Bigg\}\label{eq:GE}\\
G_{M}^{T=1}(q^{2}) & \stackrel{N_{C}\rightarrow\infty}{=} &
\frac{N_{C} M_{N}}{3} \varepsilon^{jkl} \frac{iq^{j}}{|q^{2}|}
\int d\bm{r}\: e^{i\bm{q}\cdot\bm{r}} \Bigg\{ \sum_{E_{n}\leq
E^{v}} \psi_{n}^{\dag}(\bm{r}) \gamma^{0}\gamma^{k}\tau^{l}
\psi_{n}(\bm{r})\nonumber\\ & &
-\frac{M^{2}}{M_{PV}^{2}}\sum_{E_{n}^{(PV)}\leq 0}
 \psi_{n}^{(PV)\dag}(\bm{r})
\gamma^{0}\gamma^{k}\tau^{l} \psi_{n}^{(PV)}(\bm{r})
\Bigg\}\label{eq:GM}.
\end{eqnarray}
\end{subequations}
\end{widetext}
The $\psi^{(PV)}_{n}(\bm{r})$ are the solutions of the Dirac
equation with the replacement $M\rightarrow M_{PV}$. In the
nuclear medium, Eqs.~(\ref{eq:GEM}) acquire a dependence on the
Fermi momentum $G_{X}^{T=0,1}(q^{2})\rightarrow
G_{X}^{T=0,1}(q^{2},k_{F})$ through the wave functions. This
dependence is the subject of the next section.

\section{Nuclear Physics}
\label{sec:medium}

We will begin with some motivation for our procedure to couple the
quark substructure of the nucleon to the nuclear medium. Through
the use of QCD sum rules, Ioffe \cite{Ioffe:kw} derived a
relationship between the vacuum scalar condensate, $\langle
\overline{\psi}\psi\rangle_{0}$, and the nucleon mass. One can
re-derive this estimate in a constituent quark field theory such
as we are using here. We begin with the scalar condensate
\begin{eqnarray}
\langle \overline{\psi}\psi\rangle_{0} & = & - \text{tr}
\int^{\Lambda} \frac{d^{4}p}{(2 \pi)^{4}} \frac{1}{p\!\!\!\!\;/\,
- M}
\nonumber\\
& \sim & -\frac{N_{C}M \Lambda^{2}}{4 \pi^{2}}, \label{eq:qqvac}
\end{eqnarray}
where the divergent integral is regulated by a momentum cutoff
(playing the role of the Borel mass in the QCD sum rule approach).
Using the fact that constituent quarks are essentially defined as
having a mass $\sim M_{N}/N_{C}$, we can rewrite
Eq.~(\ref{eq:qqvac}) as
\begin{eqnarray}
M_{N} & \sim & -\frac{4 \pi^{2}}{\Lambda^{2}} \langle
\overline{\psi}\psi\rangle_{0}. \label{eq:ioffe}
\end{eqnarray}
Although Eq.~(\ref{eq:ioffe}) is not a very accurate estimate, it
does highlight the role of the condensate. It will be modified in
the presence of other nucleons.

The condensate at finite density can be written in terms of the
nuclear scalar density $\rho_{s}^{N}$ and the nucleon sigma term
$\sigma_{N}$ \cite{Cohen:1991nk} as
\begin{eqnarray}
\langle \overline{\psi}\psi\rangle_{\rho} & = &
\langle\overline{\psi}\psi\rangle_{0} -
\langle\overline{\psi}\psi\rangle_{0}\frac{\sigma_{N}}{m_{\pi}^{2}f_{\pi}^{2}}
\rho_{s}^{N}\label{eq:qqmed}.
\end{eqnarray}
We can then substitute Eq.~(\ref{eq:qqmed}) into
Eq.~(\ref{eq:ioffe}) to obtain a schematic picture of the effect
of the nuclear medium on the nucleon mass
\begin{eqnarray}
M_{N}(\rho) & \sim & -\frac{4 \pi^{2}}{\Lambda^{2}} \left[
\langle\overline{\psi}\psi\rangle_{0} - c_{s} \rho_{s}^{N}
\right], \label{eq:ioffemed}
\end{eqnarray}
where $c_{s}$ is the combination of the vacuum condensate, pion
mass, decay constant and the the sigma term in
Eq.~(\ref{eq:qqmed}).

Using this dependence of the nucleon mass on the nuclear medium as
a guide, we incorporate the medium dependence in the model by
simply letting the quark scalar density in the field equation
(\ref{eq:thetafe}) contain a (constant) contribution arising from
other nucleons present in symmetric nuclear matter. This models a
scalar interaction via the exchange of multiple pairs of pions
between nucleons. We take the scalar density to consist of three
terms: 1) the constant condensate value $\langle
\overline{\psi}\psi\rangle_{0}$ (in the vacuum or at large
distances from a free nucleon), 2) the valence contribution
$\rho_{s}^{v}$ and 3) the contribution from the medium which takes
the form of the convolution of the nucleon $\rho_{s}^{N}$ and
valence quark scalar densities as in the QMC model
\cite{Lu:1998tn}
\begin{subequations}
\label{eq:rhos}
\begin{eqnarray}
\rho_{s}^{q}(\bm{r}) & \simeq & \langle
\overline{\psi}\psi\rangle_{0} + \rho_{s}^{v}(\bm{r}) +
\tilde{c}_{s} \int
d\bm{r}'\rho_{s}^{N}(\bm{r}-\bm{r}')\rho_{s}^{v}(\bm{r}')\\
& = & \langle \overline{\psi}\psi\rangle_{0} + \rho_{s}^{v}(r) +
\tilde{c}_{s} \rho_{s}^{N}S\\
& & S \equiv \int d\bm{r}' \rho_{s}^{v}(\bm{r}').
\end{eqnarray}
\end{subequations} We take the pseudoscalar density to have only
the valence term $\rho_{ps}^{q} \simeq \rho_{ps}^{v}$; the two
other contributions analogous to the first and third terms of
Eq.~(\ref{eq:rhos}) vanish due to symmetries of the QCD vacuum and
nuclear matter. These approximations to the densities neglect the
precise form of the negative continuum wave functions in
Eq.~(\ref{eq:thetafe}). The resulting free nucleon profile
function has no discernible difference from a fully
self-consistent treatment, demonstrating the excellence of this
approximation. We take $\tilde{c}_{s} = c_{s}/S$ in
Eqs.~(\ref{eq:ioffemed}) and (\ref{eq:rhos}) to be a free
parameter, which we vary to fit nuclear binding. This can be seen
as either varying $\sigma_{N}$ in Eq.~(\ref{eq:qqmed}) or the
vacuum value of the condensate in Eq.~(\ref{eq:rhos}) with
$\rho_{\Gamma}^{q}\rightarrow\rho_{\Gamma}^{q}/\tilde{c}_{s}$, as
was done in Ref.~\cite{Smith:2003hu}, since the overall
normalization cancels in Eq.~(\ref{eq:thetafe}).


The nucleon scalar density is determined by solving the nuclear
self-consistency equation
\begin{equation}
\rho_{s}^{N} = 4 \int^{k_{F}} \frac{d^{3}k}{(2\pi)^{3}}
\frac{M_{N}(\rho_{s}^{N})}{\sqrt{k^{2}+M_{N}(\rho_{s}^{N})^{2}}}.\label{eq:nsc}
\end{equation}
The dependence of the nucleon mass, and any other properties
calculable in the model, on the Fermi momentum $k_{F}$ enters
through Eq.~(\ref{eq:nsc}). Thus there are two coupled
self-consistency equations: one for the profile,
Eq.~(\ref{eq:thetafe}), and one for the density,
Eq.~(\ref{eq:nsc}). These are iterated until the change in the
nucleon mass Eq.~(\ref{eq:mn}) is as small as desired (in our
case, $\Delta M_{N}\lesssim 0.1 \text{ MeV}$) for each value of
the Fermi momentum. We use the Kahana-Ripka (KR) basis
\cite{Kahana:be}, with momentum cutoff $\Lambda$ and box size $L$
extrapolated to infinity (from a maximum value of $\Lambda L =
150$, comparable to that in Ref.~\cite{Christov:1995hr}), to
evaluate the energy eigenvalues and wave functions used as input
for the densities, nucleon mass, and electromagnetic form factors.

While the vacuum value of the condensate does not vary with the
Fermi momentum by definition, the effective condensate, $\langle
\overline{\psi}\psi\rangle_{0} + \tilde{c}_{s} \rho_{s}^{N}(k_{F})
S(k_{F})$, falls $\sim 30 \%$ at nuclear density,
\textit{q.v.}~Eq.~(\ref{eq:qqmed}). This is consistent with the
model independent result \cite{Cohen:1991nk} that predicts a value
25-50\% below the vacuum value.

A phenomenological vector meson (mass $m_{v}=770\text{ MeV}$)
exchanged between nucleons (but not quarks in the same nucleon),
is introduced as a substitute for uncalculated soliton-soliton
interactions in order to obtain the necessary short distance
repulsion which stabilizes the nucleus. This does not affect the
form factors Eqs.~(\ref{eq:GE}) and (\ref{eq:GM}). The resulting
energy per nucleon is
\begin{equation}
\frac{E}{A} = \frac{4}{\rho_{B}(k_{F})} \int^{k_{F}}
\frac{d^{3}k}{(2\pi)^{3}} \sqrt{k^{2} + M_{N}(k_{F})^{2}}
+\frac{1}{2}\frac{g_{v}^{2}}{m_{v}^{2}}\rho_{B}(k_{F})
\label{eq:epn}.
\end{equation}

\begin{figure}
\centering
\includegraphics[scale=0.5]{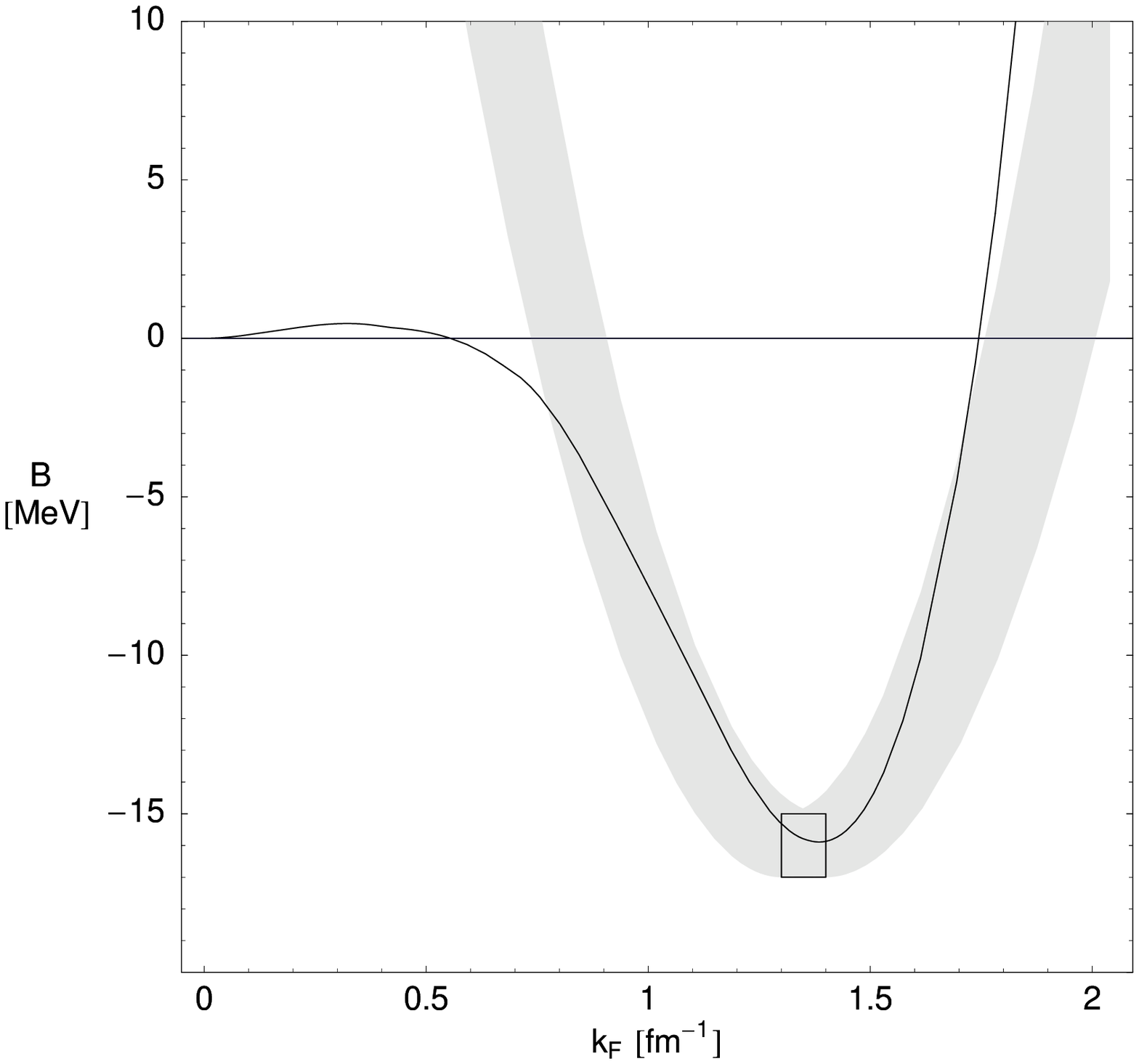}
\caption{Binding energy per nucleon $B = E/A - M_{N}$. The box and
the gray band correspond to the uncertainty in the known values of
the binding energy, density and compressibility of nuclear matter
\cite{Blaizot:tw}.} \label{fig:bepn} \centering
\includegraphics[scale=0.5]{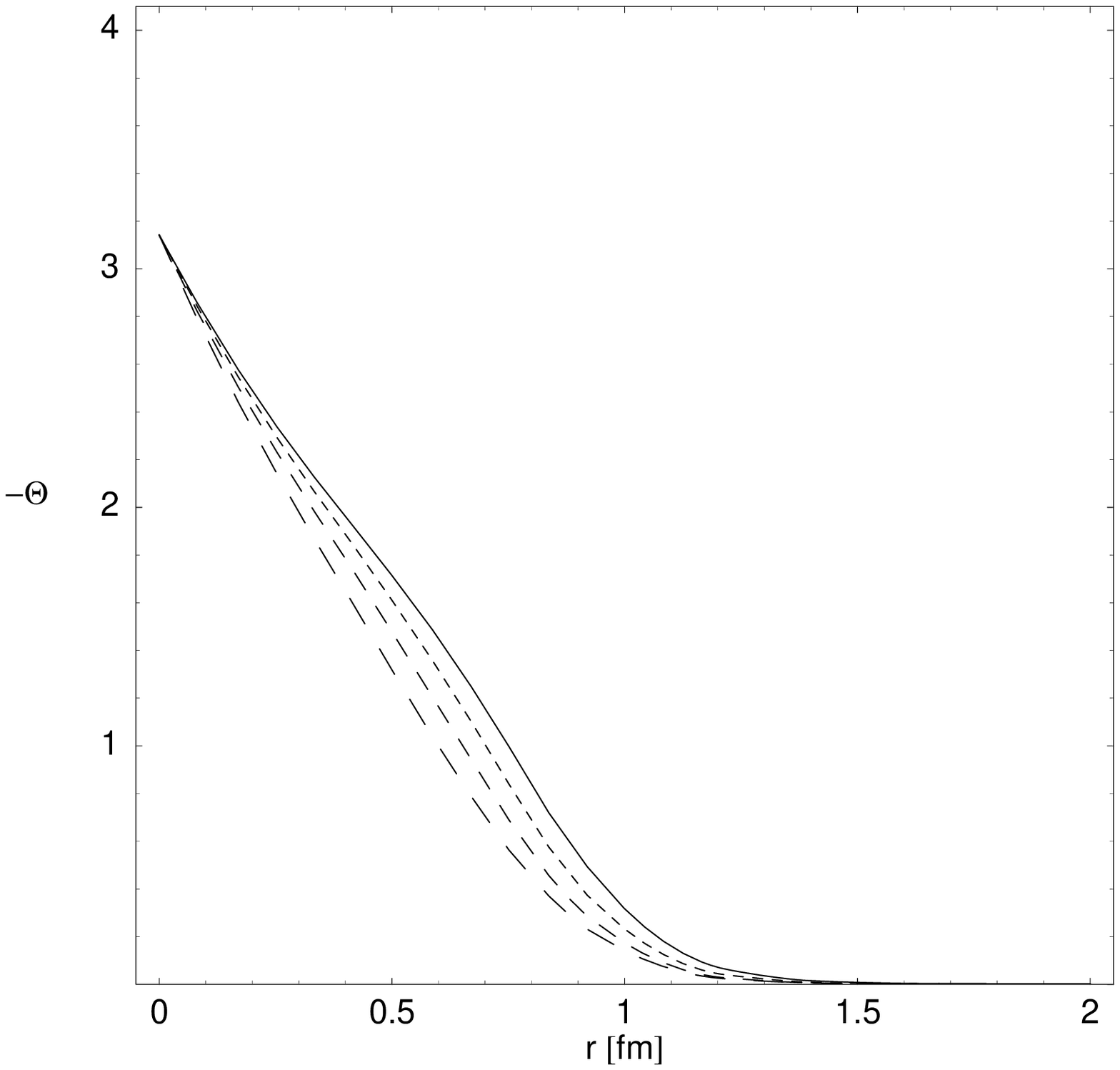}
\caption{Profile functions in nuclear matter. The solid line is
the profile function for $1.5\rho_{0}$; the curves with
progressively longer dashes correspond to $1.0\rho_{0}$,
$0.5\rho_{0}$ and zero density respectively.} \label{fig:profile}
\end{figure}

The mass of a free nucleon is computed to be
$M_{N}(k_{F}=0)=1209\text{ MeV}$. The $\sim 30\%$ difference is as
expected in the model at leading order in $N_{C}$. We evaluate the
nucleon mass Eq.~(\ref{eq:mn}) and energy per nucleon
Eq.~(\ref{eq:epn}) as a function of $k_F$. We choose our free
parameters to fit $E/A - M_{N}(0) \equiv B = -15.75 \text{ MeV}$
at the minimum. We use the value $\tilde{c}_{s} = 1.27$
(corresponding to $\sigma_{N} = 41.4\text{ MeV}$), and vector
coupling $g_{v}^{2}/4\pi = 10.55$, which gives a Fermi momentum of
$k_{F} = 1.38\text{ fm}^{-1}$ in nuclear matter consistent with
the known value $k_{F} = 1.35 \pm 0.05 \text{ fm}^{-1}$
\cite{Blaizot:tw}. We plot the binding energy per nucleon using
Eq.~(\ref{eq:epn}) in Fig.~\ref{fig:bepn}. The compressibility is
$K = 348.5\text{ MeV}$ which is above the experimental value $K =
210 \pm 30 \text{ MeV}$, but well below the Walecka model
\cite{Walecka:qa} value of $560 \text{ MeV}$. The self-consistent
calculation results in the profile functions for zero density,
$0.5\rho_{0}$, $1.0\rho_{0}$ and $1.5\rho_{0}$ in
Fig.~\ref{fig:profile} (where $\rho_{0}$ is nuclear density).

\section{Results and Discussion}
\label{sec:results}

\begin{figure}
\centering
\includegraphics[scale=0.5]{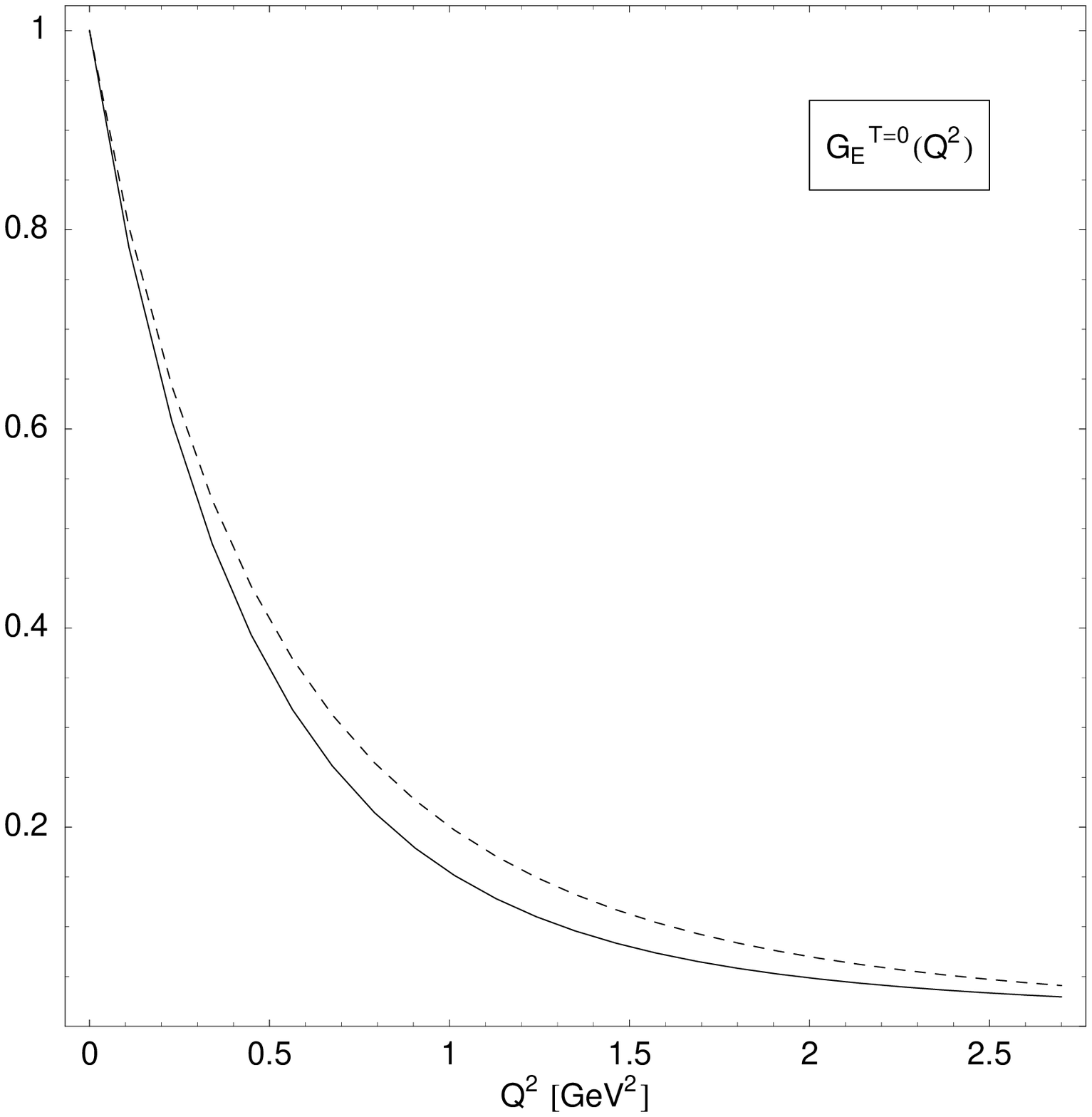}
\caption{The isoscalar electric form factor at nuclear density (solid)
and at zero density (dashes).}
\label{fig:eff} \centering
\includegraphics[scale=0.5]{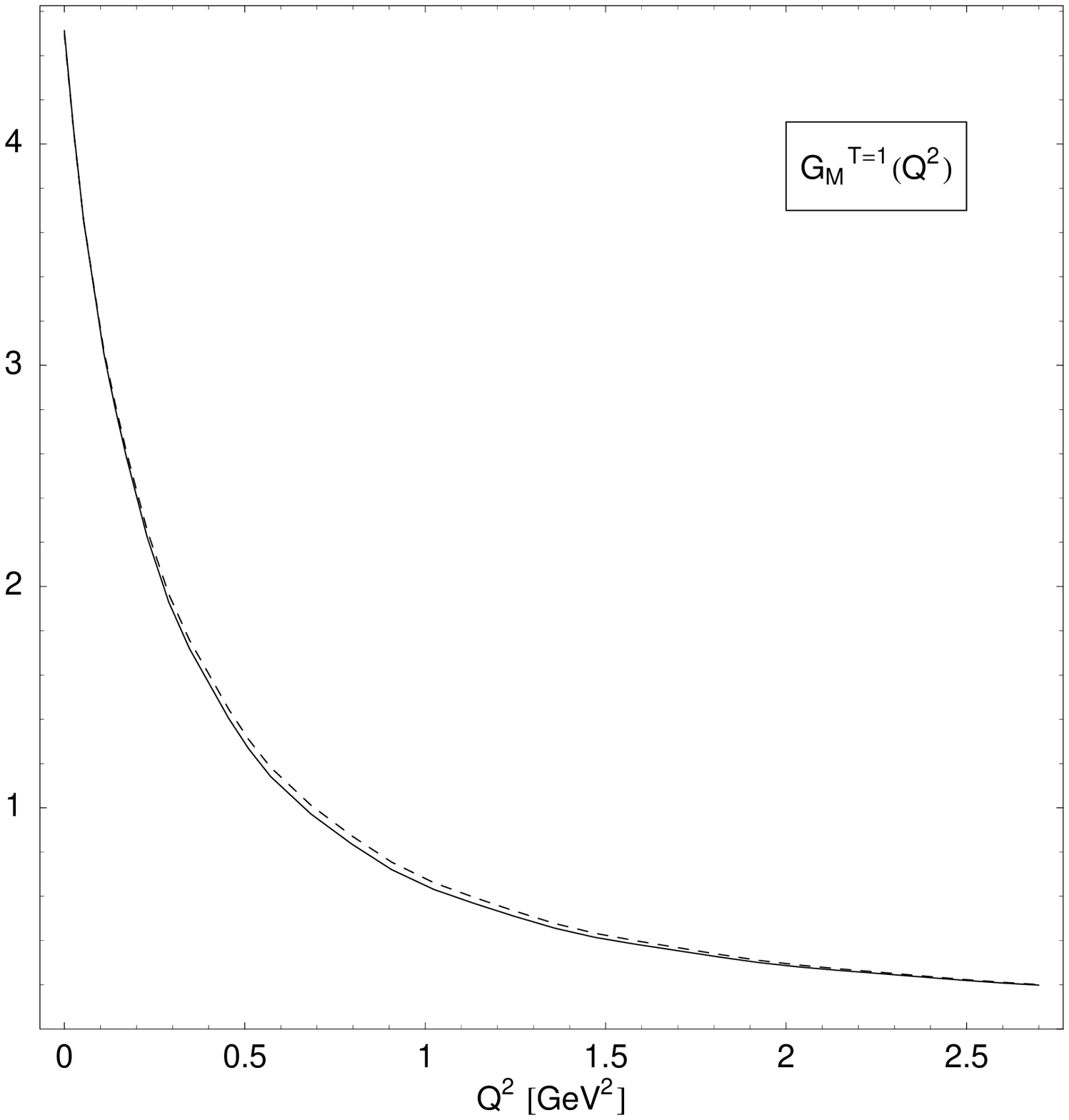}
\caption{The isovector magnetic form factor at nuclear density (solid)
and at zero density (dashes).}
\label{fig:mff}
\end{figure}

\begin{figure}
\centering
\includegraphics[scale=0.5]{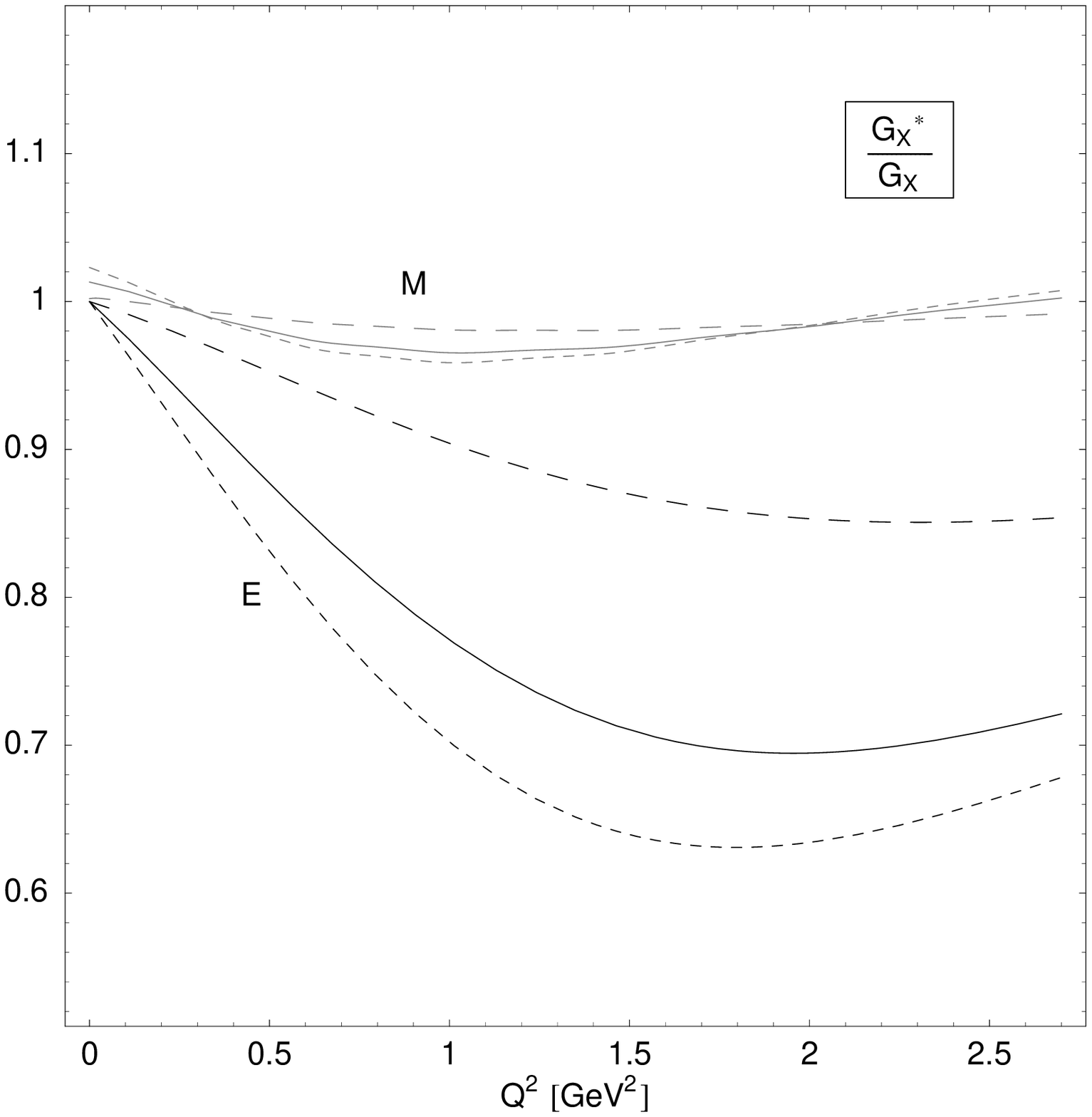}
\caption{The electric (lower three curves) and magnetic (upper
three curves) form factor ratios in Eq.~(\ref{eq:ffr}) for
$0.5\rho_{0}$ (long dashes), $1.0\rho_{0}$ (solid) and
$1.5\rho_{0}$ (short dashes).} \label{fig:ffr} \centering
\includegraphics[scale=0.5]{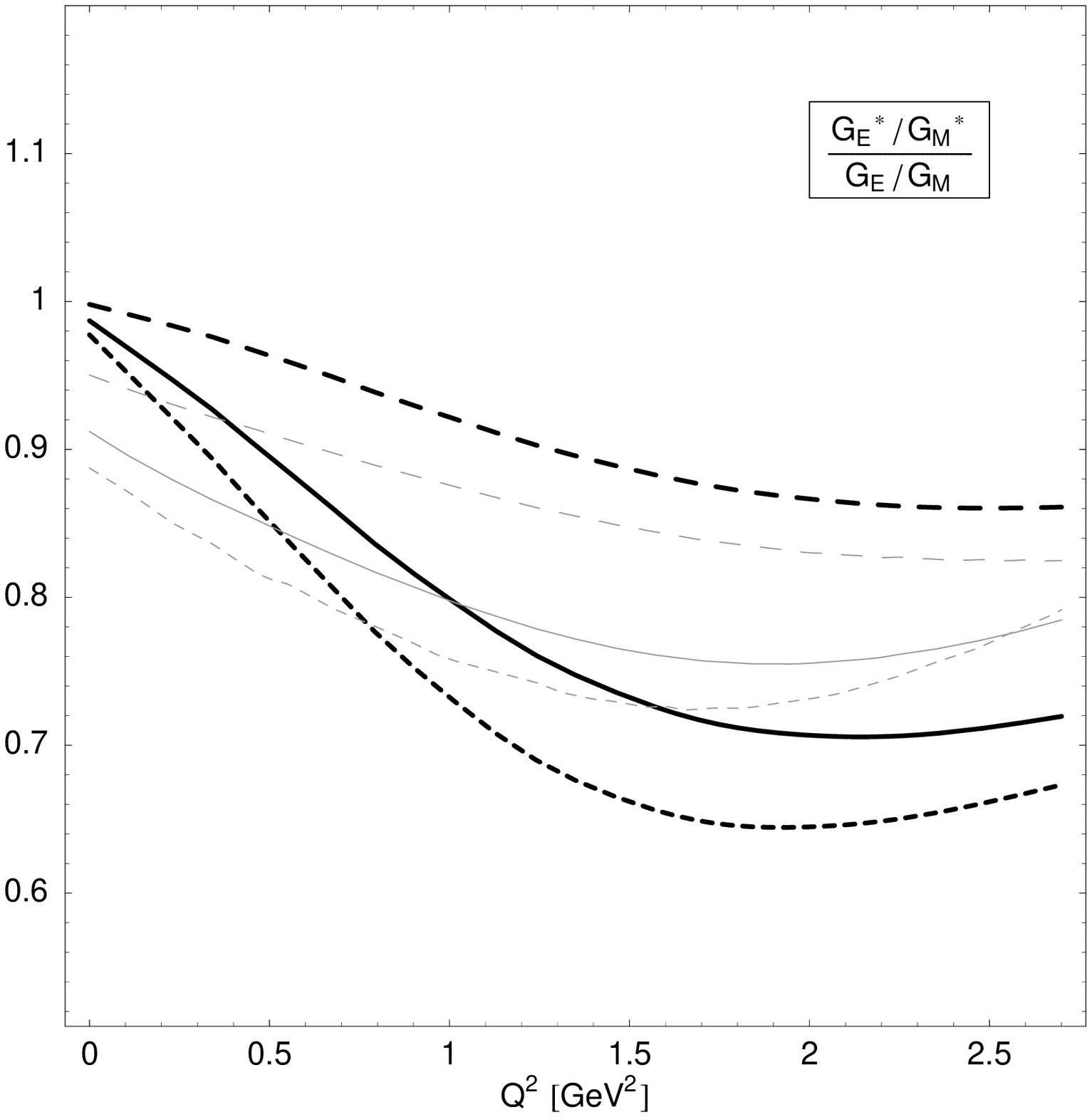}
\caption{The double ratio Eq.~(\ref{eq:ffdr}) of the electric to
magnetic form factors in nuclear matter and in the vacuum from the
CQS model (heavy) and the QMC model \cite{Lu:1998tn} (light).
Three densities are shown: $0.5\rho_{0}$ (long dashes),
$1.0\rho_{0}$ (solid) and $1.5\rho_{0}$ (short dashes).}
\label{fig:ffdr}
\end{figure}

We use Eqs.~(\ref{eq:GE}) and (\ref{eq:GM}) to calculate the form
factors, which we present in Figs.~\ref{fig:eff} and
\ref{fig:mff}.
We also present the results in terms of the ratios
\begin{equation}
\frac{G_{E,M}^{T=0,1}(Q^{2},k_{F})}{G_{E,M}^{T=0,1}(Q^{2},0)}
\equiv \frac{G_{X}^{*}(Q^{2})}{G_{X}(Q^{2})}\label{eq:ffr},
\end{equation}
where $-q^{2} \equiv Q^{2}$, $X \text{ is } E (T=0) \text{ or } M
(T=1)$, and the double ratio
\begin{equation}
\frac{G_{E}^{*}(Q^{2})/G_{M}^{*}(Q^{2})}{G_{E}(Q^{2})/G_{M}(Q^{2})}\label{eq:ffdr}.
\end{equation}
These ratios are plotted in Figs.~\ref{fig:ffr} and \ref{fig:ffdr}
for $0.5\rho_{0}$, $1.0\rho_{0}$ and $1.5\rho_{0}$.

The electric form factor is dominated by the valence contribution
and shows a dramatic effect, while the magnetic form factor has
equally important contributions from the valence and the sea. The
latter shows almost no change in nuclear matter; it shows only a
1.3\% enhancement of the magnetic moment at nuclear density, and a
2.3\% enhancement at 1.5 times nuclear density. The effect in the
electric form factor calculated here is comparable to that of the
QMC model \cite{Lu:1998tn}; the main difference from that
calculation lies in the lack of enhancement in the magnetic form
factor, specifically the practically unchanged value of the
magnetic moment.

While both form factors use the same wave functions, the isovector
magnetic form factor includes an extra weighting by a factor of
the angular momentum of the state (relative to the electric form
factor) due to the $\gamma^{k}$ in Eq.~(\ref{eq:GM}). This extra
factor is not only responsible for making the regularization of
Eq.~(\ref{eq:GM}) necessary, but for making the sea contribution
as important as the valence. In the CQS model, the orbital angular
momentum carried by the sea is comparable to the orbital angular
momentum carried by the valence quarks \cite{Wakamatsu:1990ud}
(the sum of which make up about 60\% of the total angular momentum
of the nucleon state, with the remainder belonging to the
intrinsic spin of the constituent quarks).

Conversely, the isoscalar electric form factor (which is finite,
after the vacuum subtraction) does not have as large of a
contribution from the sea. The valence level is the most important
piece, even at $Q^{2}>0$, since the $Q^{2}$ dependence in the form
factors arises from the wave functions \cite{Christov:1995hr}. The
negative Dirac continuum wave functions largely cancel in the
vacuum subtraction in Eq.~(\ref{eq:GE}).

The magnetic form factors are sensitive to the tail of the quark
wave functions, and the mere existence of a tail is due to the
lack of confinement. This is one reason for the discrepancy
between the current results and the QMC model \cite{Lu:1998tn},
but the primary source is due to the resistance to change of the
sea. The former accounts for only a few percent of the difference;
it is the latter that is our most important result. We see that
the role of antiquarks is again prevalent as in our previous work
\cite{Smith:2003hu}.

The double ratio obtained in Fig.~\ref{fig:ffdr} has the same
trend as the QMC model \cite{Lu:1998tn}, but differs in the
details. Since we obtain a similar double ratio, we expect to have
similar results if we compare these results with the polarization
transfer data \cite{Strauch:2002wu}. This requires one to take the
final state and relativistic effects into account through the use
of the RDWIA \cite{Udias:1999tm} or the RMSGA \cite{Lava:2004mp},
which accounts for a few percent of the discrepancy between the
results for bound and free protons. A RMSGA calculation for the
Helium reaction studied in Ref.~\cite{Strauch:2002wu} has been
done with these CQS model results \cite{Lava:private}, and it
delivers remarkably similar results to the same calculation done
with the QMC model \cite{Lava:2004mp}. The CQS model predicts a
smaller deviation than the QMC model from a Relativistic Plane
Wave Impulse Approximation (RPWIA) calculation, which is taken as
a baseline in Ref.~\cite{Strauch:2002wu}. While it slightly
worsens the agreement with the data at $Q^{2}\lesssim 1$, the
differences are of the same order of magnitude as the current
experimental error, and both models under predict the observed
deviation from a RPWIA calculation. At higher $Q^{2}$, the two
models produce nearly identical results for Helium.

We ignore important corrections due to the rotation of the soliton
that are suppressed by $1/N_{C}$. These corrections break the
$N-\Delta$ degeneracy, and improve the agreement of the vacuum
form factors with experiment \cite{Christov:1995hr}. More relevant
to the calculation presented here, these corrections do not affect
the $Q^2$ dependence, but instead affect the normalization of the
form factors \cite{Christov:1995hr}. However, there is no reason
at that level to continue to ignore quantum fluctuations of the
the pion field (quark loops), and treat the profile function as a
purely self-consistent mean field. We will save this difficult
problem for the future.

We have calculated the electric and magnetic form factors at
leading order in $N_{C}$ at nuclear density using the CQS model.
Our results help validate the apparent success of the QMC model in
describing the polarization transfer experiment
\cite{Strauch:2002wu,Lu:1998tn}, and provide a counterpoint to be
distinguished when finer resolution becomes available in the data.
In fact, the difference between the CQS model double ratio and the
QMC model \cite{Lu:1998tn} is roughly the size as the current
experimental error. Specifically, data on the bound nucleon
magnetic form factor at low $Q^{2}$, particularly the magnetic
moment, could serve to determine the role of sea quarks in nuclei.

\begin{acknowledgments}
We would like to thank the USDOE for partial support of this work.
We would also like to thank P.~Lava and J.~Ryckebusch for doing the
RMSGA calculation with the CQS model, and S.~Strauch for acceptance
averaging the results, so that a comparison with the data could be made.
\end{acknowledgments}

\end{document}